\documentclass[aps,prd,twocolumn,superscriptaddress,nofootinbib]{revtex4-1}

\usepackage{latexsym}
\usepackage{amsmath}
\usepackage{amssymb}
\usepackage{amsfonts}
\usepackage{bm}
\RequirePackage{lineno}

\usepackage{color}
\definecolor{purple}{rgb}{0.5,0,0.5}
\definecolor{blue}{rgb}{0.0,0,0.9}
\definecolor{prdblue}{rgb}{0.133,0.118,0.498}
\usepackage[colorlinks=true, pdfstartview=FitV, linkcolor=prdblue, citecolor= prdblue, urlcolor=prdblue]{hyperref}

\usepackage{supertabular} 
\usepackage{placeins}
\usepackage{epsfig}
\usepackage{graphicx}
\usepackage{hyperref}

\hypersetup{
  breaklinks=true,
  colorlinks=true,
  linkcolor=blue,
  anchorcolor=blue,
  citecolor=blue,
  filecolor=blue,
  urlcolor=blue
}
\hypersetup{
  bookmarksnumbered=true,
  bookmarkstype=toc,
  linktocpage=true
}

\begin{document}

\modulolinenumbers[2]

\setlength{\oddsidemargin}{-0.5cm} \addtolength{\topmargin}{15mm}

\title{\boldmath Precise measurement of the form factors in $D^0\rightarrow K^*(892)^-\mu^+\nu_{\mu}$ and test of lepton universality with $D^0\rightarrow K^*(892)^-\ell^+\nu_{\ell}$ decays }

\author{
  \small
M.~Ablikim$^{1}$, M.~N.~Achasov$^{4,c}$, P.~Adlarson$^{77}$, X.~C.~Ai$^{82}$, R.~Aliberti$^{36}$, A.~Amoroso$^{76A,76C}$, Q.~An$^{73,59,a}$, Y.~Bai$^{58}$, O.~Bakina$^{37}$, Y.~Ban$^{47,h}$
% ...
 \\
 \vspace{0.2cm}
 (BESIII Collaboration)\\
 \vspace{0.2cm} {\it
$^{1}$ Institute of High Energy Physics, Beijing 100049, People's Republic of China\\
% ...
$^{4}$ Budker Institute of Nuclear Physics SB RAS (BINP), Novosibirsk 630090, Russia\\
% ...
$^{36}$ Johannes Gutenberg University of Mainz, Johann-Joachim-Becher-Weg 45, D-55099 Mainz, Germany\\
$^{37}$ Joint Institute for Nuclear Research, 141980 Dubna, Moscow region, Russia\\
% ...
$^{47}$ Peking University, Beijing 100871, People's Republic of China\\
% ...
$^{58}$ Southeast University, Nanjing 211100, People's Republic of China\\
$^{59}$ State Key Laboratory of Particle Detection and Electronics, Beijing 100049, Hefei 230026, People's Republic of China\\
% ...
$^{73}$ University of Science and Technology of China, Hefei 230026, People's Republic of China\\
% ...
$^{76}$ University of Turin and INFN, (A)University of Turin, I-10125, Turin, Italy; (B)University of Eastern Piedmont, I-15121, Alessandria, Italy; (C)INFN, I-10125, Turin, Italy\\
$^{77}$ Uppsala University, Box 516, SE-75120 Uppsala, Sweden\\
% ...
$^{82}$ Zhengzhou University, Zhengzhou 450001, People's Republic of China\\
\vspace{0.2cm}
$^{a}$ Deceased\\
% ...
$^{c}$ Also at the Novosibirsk State University, Novosibirsk, 630090, Russia\\
% ...
$^{h}$ Also at State Key Laboratory of Nuclear Physics and Technology, Peking University, Beijing 100871, People's Republic of China\\
% ...
\vspace{0.4cm}
}
}
\begin{abstract}
We report a study of the semileptonic decay $D^0 \rightarrow \bar{K}^0\pi^-\mu^+\nu_{\mu}$ based on a sample of $7.9~\mathrm{fb}^{-1}$ of $e^+e^-$ annihilation data collected at a center-of-mass energy of 3.773~GeV with the BESIII detector at the BEPCII collider. The branching fraction of the decay is measured for the first time to be $\mathcal{B}(D^0\rightarrow \bar{K}^0\pi^-\mu^+\nu_{\mu}) = (1.373 \pm 0.020_{\rm stat} \pm 0.023_{\rm syst})\%$, where the first uncertainty is statistical and the second is systematic. Based on the investigation of the decay dynamics, we find that the decay is dominated by the $K^{*}(892)^-$ resonance with the branching fraction measured to be $\mathcal{B}(D^0\rightarrow K^{*}(892)^-\mu^+\nu_{\mu}) = (1.948 \pm 0.033_{\rm stat} \pm 0.036_{\rm syst})\%$. We also determine the hadronic form factors for the $D^0\rightarrow K^{*}(892)^-\mu^+\nu_{\mu}$ decay to be $r_{V} = V(0)/A_1(0) = 1.46 \pm 0.11_{\rm stat} \pm 0.04_{\rm syst}$, $r_{2} = A_2(0)/A_1(0) = 0.71 \pm 0.08_{\rm stat} \pm 0.03_{\rm syst}$, and $A_1(0)=0.609 \pm 0.008_{\rm stat} \pm 0.008_{\rm syst}$, where $V(0)$ is the vector form factor and $A_{1,2}(0)$ are the axial form factors evaluated at $q^2=0$. The $A_1(0)$ is measured for the first time in $D^0\rightarrow K^{*}(892)^-\mu^+\nu_{\mu}$ decay. Averaging the form-factor parameters that we reported previously in $D^0\rightarrow K^*(892)^-(\rightarrow \bar{K}^0\pi^-)e^+\nu_{e}$ and $D^0\rightarrow K^*(892)^-(\rightarrow K^-\pi^0)\mu^+\nu_{\mu}$ decays, we obtain $r_{V}=1.456\pm0.040_{\rm stat}\pm0.016_{\rm syst}$, $r_{2}=0.715\pm0.031_{\rm stat}\pm0.014_{\rm stat}$, and $A_1(0)=0.614\pm0.005_{\rm stat}\pm0.004_{\rm syst}$. This is the most precise determination of the form-factor parameters to date measured in $D\rightarrow K^*(892)$ transition, which provide the most stringent test on various theoretical models.
\end{abstract}


\maketitle

Semileptonic (SL) decays of $D$ mesons provide valuable information on the weak and strong interactions within hadrons composed of heavy quarks~\cite{physrept494,RevModPhys67_893}.
The partial decay rate is related to the product of the hadronic form factor (FF) describing the effects of the strong interaction in the initial and final hadrons, and the Cabibbo-Kobayashi-Maskawa (CKM) matrix~\cite{prl10_531} element $|V_{cs(d)}|$ parametrizing the mixing of quarks in the weak interaction. 
Due to $|V_{cs(d)}|$ being tightly constrained by the CKM unitarity, studies of $D$ meson SL decays provide an ideal laboratory to extract the hadronic FFs. 
In recent years, there has been an accumulation of hadronic FF measurements in $D\rightarrow V\ell^+\nu_{\ell}$ decays~\cite{prl122_061801,prd78_051101,prd74_052001,prd81_112001,prd83_072001,prd94_032001,plb607_67,prd99_011103,D0Kspiev,2403.10877,prl110_131802,prd92_071101}, where $V$ and $\ell$ refer to a vector meson and a lepton, respectively. In particular, the SL decay $D^0\rightarrow K^{*}(892)^-\ell^+\nu_{\ell}$ plays a crucial role in understanding the standard model (SM), as it provides the most direct way to experimentally extract the value of $|V_{cs}|$. However, such an extraction requires precise theoretical knowledge of the hadronic FFs taking into account non-perturbative quantum chromodynamic (QCD) effects.

In theory, although the study of the $D\rightarrow K^*(892)$ transition is very challenging, calculations are extensively carried out by a series of non-perturbative approaches including Lattice QCD~\cite{Lattice}, light-cone sum rules (LCSR)~\cite{IJMPA21_6125,PRR2}, the constituent quark model (CQM)~\cite{PRD62_014006,PRD96_016017}, heavy quark effective theory (HQEFT)~\cite{PRD67_014024}, the combined heavy meson and chiral theory (HM$\chi$T)~\cite{PRD72_034029}, the covariant light-front quark model (CLFQM)~\cite{JPG39_025005,EPJC77_587}, the large energy chiral quark model (LE$\chi$QM)~\cite{PRD89_034013}, the covariant confined quark model (CCQM)~\cite{FrontPhys14_64401}, the relativistic quark model (RQM)~\cite{prd101_013004}, and holographic QCD~\cite{prd109_026008}. 
However, the FF ratios of $r_V$, $r_2$, and $A_1(0)$ in $D\rightarrow K^*(892)$ transition predicted by these  models~\cite{Lattice,IJMPA21_6125,PRR2,PRD62_014006,PRD96_016017,PRD67_014024,PRD72_034029,JPG39_025005,EPJC77_587,PRD89_034013,FrontPhys14_64401,prd101_013004,prd109_026008} differ significantly, and vary in the ranges from $1.36-1.60$, $0.50-0.92$, and $0.57-0.66$, respectively. 
Currently, the precision in measuring $r_V$ and $r_2$ is still poor~\cite{pdg24} and no direct measurement of $A_1(0)$ is available in $D^0\rightarrow K^{*}(892)^-\mu^+\nu_{\mu}$ decay~\cite{pdg24}. Therefore, precise measurements on these FF parameters can provide stringent tests and rigorous calibrations on various non-perturbative calculations. 
Furthermore, these data also represent a valuable spur for updating Lattice QCD calculations~\cite{Lattice}. 

In the SM, the SL decays of $D$ mesons offer an excellent opportunity to test lepton flavor universality (LFU)~\cite{ARNPS,NSR}. In Refs.~\cite{PRD91_094009,CPC45_063107,PRD96_016017,PRR2}, it is suggested that observable LFU violation effects may occur in SL decays via $c\rightarrow s\ell^+\nu_{\ell}$. In particular, the multiple polarization states of vector mesons in $D\rightarrow V\ell^+\nu_{\ell}$ decay provide more physical information,
allowing a more detailed search for potential new physics effects.
In recent years, the branching fraction (BF) of $D^0\rightarrow K^*(892)^-\mu^+\nu_{\mu}$ and the ratio $R^{\mu/e}_{K^*(892)}=\frac{\mathcal{B}(D^0\rightarrow K^*(892)^-\mu^+\nu_{\mu})}{\mathcal{B}(D^0\rightarrow K^*(892)^-e^+\nu_{e})}$ have been calculated using various theoretical models, leading to differing expectations. The predicted BF varies by $(1.82-3.09)\%$~\cite{PRD62_014006,PRD96_016017,PRD67_014024,PRD72_034029,IJMPA21_6125,JPG39_025005,EPJC77_587,FrontPhys14_64401,PRD89_034013,prd101_013004,PRR2}, and $R^{\mu/e}_{K^*(892)}$ is found to be between $0.92-0.95$ in Refs.~\cite{IJMPA21_6125,PRD96_016017,EPJC77_587,FrontPhys14_64401,PRD89_034013,prd101_013004,PRD92_054038} and $0.99$ in Ref.~\cite{PRR2}. 
Furthermore, the ratio of the decay rates ($R_{L/T}$) between the longitudinally and transversely polarized $K^*$ fractions in $D^0\rightarrow K^*(892)^-\mu^+\nu_{\mu}$ decay is predicted to be sensitive to the pseudoscalar Wilson coefficient $c_P^{(\ell)}$~\cite{PRD91_094009}, while no corrresponding measurement has been reported yet~\cite{pdg24}.

The SL decay $D^0\rightarrow \bar{K}^0\pi^-\mu^+ \nu_{\mu}$ was first measured by the FOCUS collaboration~\cite{plb607_67}, where only the FF ratios $r_V$ and $r_2$ were reported with limited precision. No other measurements have been performed in the last two decades.
In this Letter, we report the first measurement of the absolute BF for $D^0\rightarrow \bar{K}^0\pi^-\mu^+\nu_{\mu}$, and the most precise measurements of FF parameters $r_V$, $r_2$, and $A_1(0)$ in the decay $D^0\rightarrow K^*(892)^-\ell^+\nu_{\ell}$. Throughout this Letter, charge-conjugate modes are implied unless explicitely noted.
These measurements are performed using a data sample corresponding to an integrated luminosity of $7.9~\mathrm{fb}^{-1}$ produced at the center-of-mass energy $\sqrt{s}=3.773$~GeV with the BEPCII $e^+e^-$ collider and collected by the BESIII detector~\cite{Ablikim:2009aa}.

Simulated data samples produced with a {\sc geant4}-based Monte Carlo (MC) package~\cite{geant4}, which includes the geometric description of the BESIII detector and the detector response,
are used to determine detection efficiencies and to estimate background contributions. The simulation
models the beam energy spread and initial state radiation (ISR) in the $e^+e^-$ annihilations
with the {\sc kkmc} generator~\cite{kkmc}.
The inclusive MC sample includes the production of $D\bar{D}$ pairs, the non-$D\bar{D}$ decays of the $\psi(3770)$, the ISR
production of the $J/\psi$ and $\psi(3686)$ states, and the continuum processes incorporated in {\sc kkmc}~\cite{kkmc}. 
All particle decays are modeled with {\sc evtgen}~\cite{nima462_152} using branching fractions either taken from
the Particle Data Group~\cite{pdg24}, when available, or otherwise estimated with {\sc lundcharm}~\cite{lundcharm}.
Final state radiation (FSR) from charged final state particles is incorporated using the
{\sc photos} package~\cite{plb303_163}. The generation of the signal $D^0\rightarrow \bar{K}^0\pi^-\mu^+\nu_{\mu}$ incorporates knowledge of the FFs obtained in this work. 

Our analysis makes use of both ``single-tag'' (ST) and ``double-tag'' (DT) samples of $D$ decays.
The ST samples are $\bar{D}^0$ events reconstructed from one of their six hadronic decays as listed in Table~\ref{tab:numST}, while the DT samples are events with a ST and a $D^0$ meson reconstructed as $D^0\rightarrow \bar{K}^0\pi^-\mu^+\nu_{\mu}$.  
The BF for the SL decay is given by~\cite{D0Kspiev}
\begin{equation}
  \mathcal{B}_{\rm SL} \,=\,
  \frac{N_{\rm DT}}{\sum_i N^{i}_{\rm ST} \,
      \left(\epsilon^i_{\rm DT}/\epsilon^i_{\rm ST}\right)} \,=\,
  \frac{N_{\rm DT}}{N_{\rm ST} \, \epsilon_{\rm SL}}, \label{eq:branch}
\end{equation}
where $N_{\rm DT(ST)}$ is the total yield of DT(ST) events, $\epsilon_{\rm SL}=(\sum_i N^{i}_{\rm ST}\times\epsilon^i_{\rm DT}/\epsilon^i_{\rm ST})/\sum_i N^{i}_{\rm ST}$ is the average efficiency of reconstructing the SL decay in a ST event,
weighted by the measured yields of ST modes in the data, and $\epsilon^i_{\rm ST}$ and $\epsilon^i_{\rm DT}$ are the efficiencies for finding the ST and the SL decay in the $i$-th tag mode, respectively.

\begin{figure}[tp!]
\begin{center}
\includegraphics[width=\linewidth]{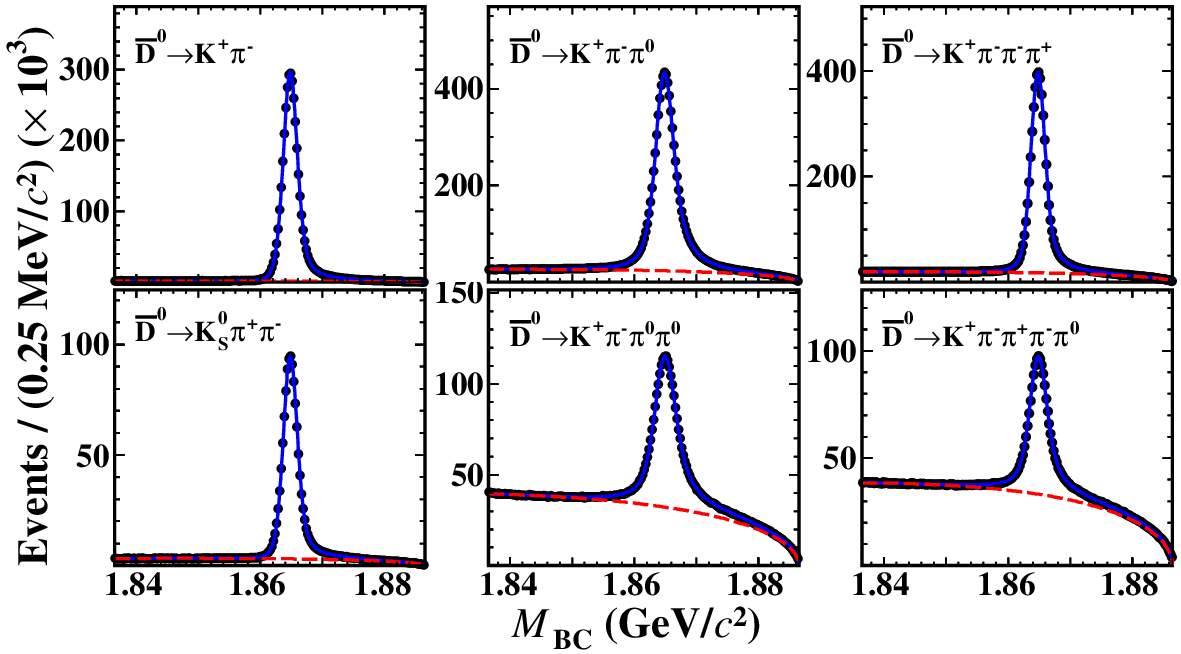}
\caption{(Color online)~The $M_{\rm BC}$ distributions of the six ST modes. The points are data, the solid blue curves are the projection of the sum of all fit components and the dashed red curves are the projection of the background component of the fit. }
\label{fig:tag_md0}
\end{center}
\end{figure}

A detailed description of the selection criteria for $\pi^{\pm}$, $K^{\pm}$, $\gamma$, $\pi^0$, $K^0_S$, and $\pi^0$ candidates is given in Ref.~\cite{D0Kspiev}.
The ST $\bar{D}^0$ mesons are identified using the beam constrained mass: 
\begin{equation}
M_{\rm BC} = \sqrt{(\sqrt{s}/2)^2-|\vec {p}_{\bar D^0}|^2},
\end{equation}
where $\vec {p}_{\bar D^0}$ is the momentum of the $\bar{D}^0$ candidate in the rest frame of the initial $e^+e^-$ system. 
A kinematic variable $\Delta E = E_{\bar{D}^0} -\sqrt{s}/2$ for
each candidate is used to improve the signal significance for ST $\bar{D}^0$ mesons, where $E_{\bar{D}^0}$ is the energy of the $\bar{D}^0$ candidate. 
The explicit $\Delta E$ requirements for these six ST modes are listed in Table~\ref{tab:numST}. 
The distributions of the variable $M_{\rm BC}$ for the six ST
modes are shown in Fig.~\ref{fig:tag_md0}, where maximum likelihood fits to the $M_{\rm BC}$ distributions are performed. 
The signal function is derived from the convolution of the MC-simulated signal shape with a double-Gaussian function to account for resolution difference between simulation and data, where the parameters of the double-Gaussian function are floated. An ARGUS function~\cite{plb241_278} is used to describe the combinatorial background shape.
For each tag mode, the ST yield is obtained by integrating the signal
function over the $D^0$ signal region within $1.859<M_{\rm BC}<1.873$~GeV/$c^2$.
The ST yields and the ST efficiencies for six ST modes are listed in Table~\ref{tab:numST}. The total ST yield summed over all six ST modes is 
$N_{\rm ST}=(7895.8\pm3.4)\times 10^3$, where the uncertainty is statistical only.

\begin{table}[tp!]
\caption{ The selection requirements on $\Delta E$, the ST yields $N_{\rm ST}$ in data, the ST and DT efficiencies, $\epsilon_{\rm ST}$ and $\epsilon_{\rm DT}$, for each of the six tag decay modes. For $\epsilon_{\rm DT}$, the BF of $\bar{K}^0\rightarrow \pi^+\pi^-$ is not included.  }
\begin{center}
\resizebox{!}{1.6cm}{
\begin{tabular}
{lcccc} \hline\hline ST mode  & $\Delta E$ (GeV)     &  $N^i_{\rm ST}$ ($\times 10^3$)   & $\epsilon^i_{\rm ST}$  (\%)    & $\epsilon^i_{\rm DT}$  (\%)   \\
\hline $K^+\pi^-$                   & [$-$0.027, 0.027]          &  $1449.3\pm1.3$ & $65.34(01)$     &   $13.88(02)$    \\
       $K^+\pi^-\pi^-\pi^+$       & [$-$0.026, 0.024]          &  $1944.2\pm1.6$ & $40.83(01)$     &   ~$6.96(02)$   \\
       $K^+\pi^-\pi^0$              & [$-$0.062, 0.049]          &  $2913.2\pm2.0$  & $35.59(01)$    &  ~$7.15(02)$    \\
       $K_S^0\pi^+\pi^-$         & [$-$0.024, 0.024]          &   ~$447.6\pm0.7$ & $37.49(01)$     &  ~$6.05(02)$    \\
       $K^+\pi^-\pi^0\pi^0$      & [$-$0.068, 0.053]          &   ~$690.6\pm1.3$ & $14.83(01)$     &  ~$2.89(01)$   \\
       $K^+\pi^-\pi^+\pi^-\pi^0$ & [$-$0.057, 0.051]        &  ~$450.9\pm1.1$  & $16.17(01)$    &   ~$2.40(01)$    \\       
\hline\hline
\end{tabular}
}
\label{tab:numST}
\end{center}
\end{table}

Candidates for the SL decay $D^0\rightarrow \bar{K}^0\pi^-\mu^+\nu_{\mu}$ are selected from the remaining tracks recoiling against the ST $\bar{D}^0$ mesons. The $\bar{K}^0$ meson is reconstructed as a $K^0_S \to \pi^+\pi^-$ decay with the same selection criteria used in the ST side. 
In addition to the two charged tracks originating from $K^0_S$ decays, the event must contain exactly two additional tracks with opposite charges.
The track having the same charge of the kaon on the tagging side is taken as the muon candidate.
For muon particle identification (PID), the ${\rm d}E/{\rm d}x$ and time-of-flight system (TOF) measurements are combined with shower properties from the electromagnetic calorimemeter (EMC) to construct likelihoods for electron, muon, pion, and kaon hypotheses, $\mathcal{L}_e$, $\mathcal{L}_{\mu}$, $\mathcal{L}_\pi$, and $\mathcal{L}_K$. The muon candidate must satisfy $\mathcal{L}_{\mu} > 0.001$, $\mathcal{L}_{\mu}>\mathcal{L}_e$, $\mathcal{L}_{\mu}>\mathcal{L}_{K}$, and $\mathcal{L}_{\mu}/(\mathcal{L}_{\mu}+\mathcal{L}_{\pi}+\mathcal{L}_K)>0.3$.
Additionally, the deposited EMC energy of the muon candidate (${\rm EMC}_{\mu}$) should be within $0.1<{\rm EMC}_{\mu}<0.3$~GeV.
The other charged track is taken as the pion candidate and must satisfy $\mathcal{L}^{\prime}_\pi>\mathcal{L}^{\prime}_K$, where $\mathcal{L}^{\prime}_\pi$ and $\mathcal{L}^{\prime}_K$ are PID likelihoods calculated with ${\rm d}E/{\rm d}x$ and TOF measurements for the pion and kaon hypotheses, respectively. For the ST mode $K_S^0\pi^+\pi^-$, it is possible that two combinations are selected for the muon and pion candidates. Only the combination with the maximum product likelihoods $\mathcal{L}_{\mu}\mathcal{L}^{\prime}_\pi$ is retained. The background candidates from $D^0\rightarrow \bar{K}^0\pi^+\pi^-(\pi^0)$ decays reconstructed as $D^0\rightarrow \bar{K}^0\pi^-\mu^+\nu_{\mu}$ are rejected by requiring the $\bar{K}^0\pi^-\mu^+(\pi^0)$ invariant mass ($M_{\bar K^0\pi^-\mu^+(\pi^0)}$) to be less than 1.60~GeV/$c^2$, where the $\pi^0$ is reconstructed with the residual photon pairs in the signal side. Background events containing additional $\pi^0$ mesons are further suppressed by requiring the maximum energy of any unused photon ($E_{\gamma \rm max}$) to be less than 0.15~GeV.

\begin{figure}[tp!]
\begin{center}
   \includegraphics[width=\linewidth]{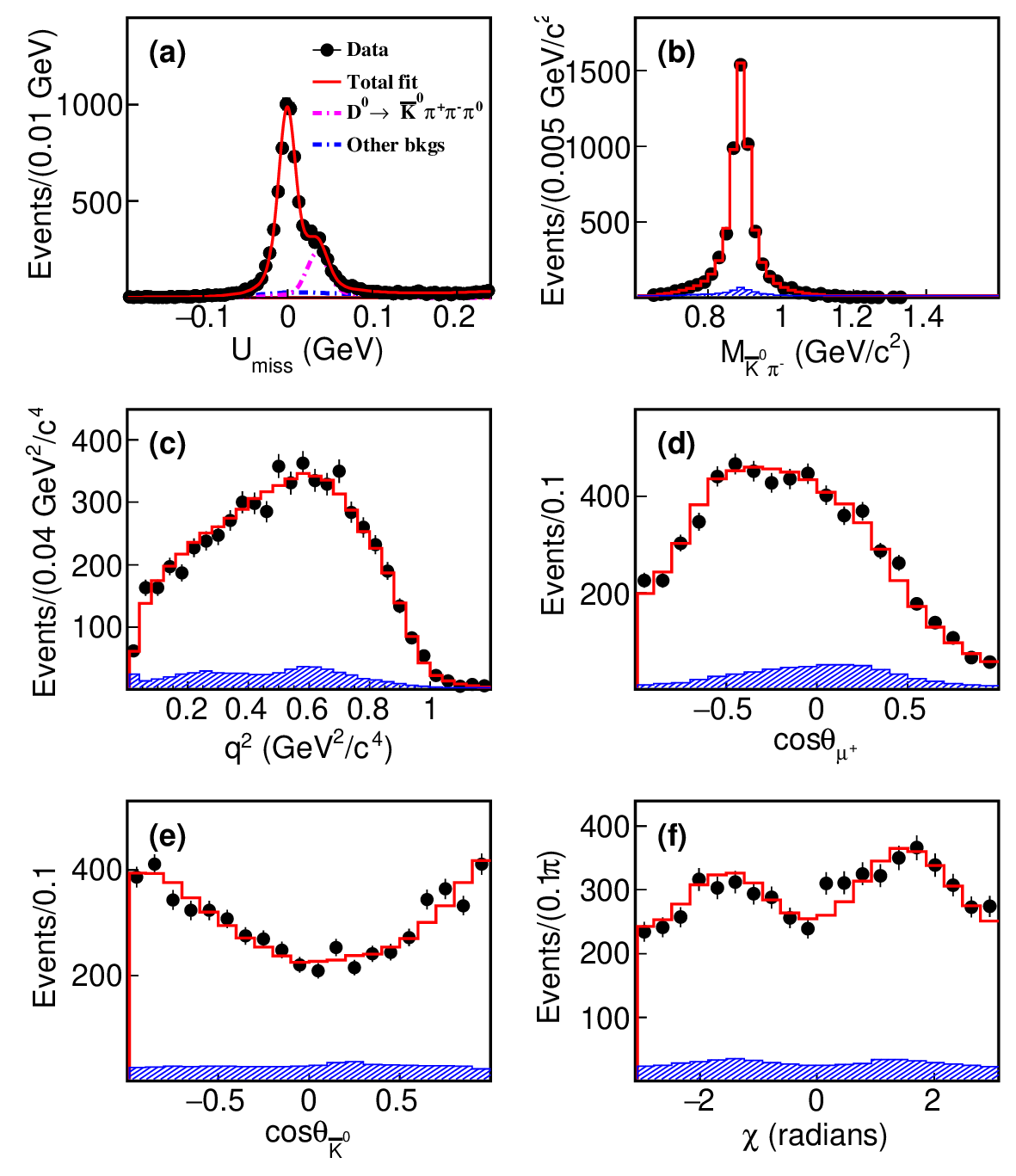}
   \caption{ (Color online)~(a) Fit of the SL candidate events to the $U_{\rm miss}$ distribution. Distributions of the five kinematic variables (b) $M_{\bar{K}^0\pi^-}$, (c) $q^2$, (d) $\cos\theta_{\mu^+}$, (e) $\cos\theta_{\bar{K}^0}$, and (f) $\chi$ for the SL decay $D^0\rightarrow \bar{K}^0\pi^-\mu^+\nu_{\mu}$. The dots with error bars are data, the red curves and histograms are the fit results, and the shaded histograms are the simulated background.}
\label{fig:formfactor}
\end{center}
\end{figure}

As the neutrino is not detected, we employ the kinematic variable
$U_{\rm miss}=E_{\rm miss}-c|\vec{p}_{\rm miss}|$ to obtain information on the neutrino, where $E_{\rm miss}$ and $\vec{p}_{\rm miss}$ are the missing energy and momentum carried by the neutrino, respectively, and are defined in the same way as in Ref.~\cite{D0Kspiev}.
Figure~\ref{fig:formfactor}(a) shows the $U_{\rm miss}$ distribution of the accepted candidates for $D^0\rightarrow \bar{K}^{0}\pi^-\mu^+\nu_{\mu}$ in data.
To obtain the signal yield, an unbinned maximum likelihood fit to the $U_{\rm miss}$ distribution is performed. 
In the fit, the signal is described with a shape derived from the simulated signal events convolved with a Gaussian function, where the mean and width of the Gaussian function are determined by the fit. The peaking background from $D^0\rightarrow \bar{K}^0\pi^-\pi^+\pi^0$ is modeled using the MC-derived shape. The other combinatorial background contribution is described using the shape obtained from the inclusive MC simulation.
The yield of $D^0\rightarrow \bar{K}^0\pi^-\mu^+\nu_{\mu}$ events is $N_{\rm DT}=6796\pm98$, where the uncertainty is statistical only.

The DT efficiency $\epsilon^i_{\rm DT}$ for each tag mode is summarized in the last column of Table~\ref{tab:numST}, and the average efficiency of reconstructing SL decay, $\epsilon_{\rm SL}$, is estimated to be $(18.98\pm0.01)\%$. The difference of the $K^0_S$ reconstruction efficiencies between data and MC is estimated to be $-(1.8\pm0.6)\%$, taking into account the systematic uncertainties due to the tracking efficiencies for the charged pions, the $K^0_S$ mass window and decay length requirements.
The difference due to the $\mu$ PID efficiency between data and MC is estimated to be $-(2.8\pm0.4)\%$ evaluated using the control sample of 
$e^+e^-\rightarrow \gamma\mu^+\mu^-$ decay.
Hence, $\epsilon_{\rm SL}$ is further corrected by $-1.8\%$ and $-2.8\%$, giving $(18.12\pm0.01)\%$~\cite{BFK0}.
The BF of the decay is measured to be $\mathcal B({D^0\rightarrow \bar{K}^{0}\pi^-\mu^+\nu_{\mu}})=(1.373\pm0.020_{\rm stat})\%$. 

Due to the DT technique, the BF measurement is insensitive to the systematic uncertainty in the ST efficiency.
The uncertainty on the muon tracking (PID) efficiency is estimated to be 0.3\%~(0.4\%) by studying a sample of $e^+e^-\rightarrow \gamma \mu^+\mu^-$ events.
The uncertainty due to the pion tracking (PID) efficiency is estimated to be 0.3\%~(0.3\%) using control samples selected from $D^0\rightarrow K^-\pi^+(\pi^0, \pi^+\pi^-)$, and $D^+\rightarrow K^-\pi^+\pi^+(\pi^0)$. The uncertainty from $K^0_S$ reconstruction is 0.6\%, determined with control samples selected from
$D^0 \rightarrow \bar{K}^0\pi^+\pi^-, \bar{K}^0\pi^+\pi^-\pi^0, \bar{K}^0\pi^0$, and  $D^+ \rightarrow \bar{K}^0\pi^+, \bar{K}^0\pi^+\pi^0, \bar{K}^0\pi^+\pi^+\pi^-$.
 The uncertainty associated with the $E_{\gamma\,{\rm \max}}$ requirement is estimated to be 0.7\% by analyzing DT $D^0\bar{D}^0$ events where $D^0$ mesons decay to hadronic final states of $D^0\rightarrow K^-\pi^+$, $K^-\pi^+\pi^0$, and $K^-\pi^+\pi^+\pi^-$. The uncertainty due to the $M_{\bar K^0\pi^-\mu^+(\pi^0)}$ requirement is estimated to be 0.9\% evaluated using a control sample from $D^+\rightarrow K^-\pi^+e^+\nu_{e}$ with the positron mass substituted by the $\mu$ mass. The uncertainty due to the modeling of the signal in simulated events is estimated to be 0.7\% by varying the input FF parameters determined in this work by $\pm 1\sigma$. The uncertainty associated with the fit to the $U_{\rm miss}$ distribution is estimated to be 0.6\% by varying the fitting ranges and the shapes which parametrize the signal and background candidates, where an asymmetric Gaussian function is used as an alternative signal function.
The uncertainty associated with the fit of the $M_{\rm BC}$ distributions used to determine $N_{\rm ST}$ is 0.1\% and is evaluated by varying the bin size, fit range and background distributions. Further systematic uncertainties are assigned due to the statistical precision of the simulation, 0.3\%, and the input BF of the decay $K^0_S\rightarrow \pi^+ \pi^-$, 0.1\%. The systematic uncertainty contributions are summed in quadrature, and the total systematic uncertainty on the BF measurement is 1.7\%. Finally, we obtain $\mathcal B({D^0\rightarrow \bar{K}^{0}\pi^-\mu^+\nu_{\mu}})=(1.373\pm0.020_{\rm stat}\pm 0.023_{\rm syst})\%$.

The differential decay rate of $D^0\rightarrow \bar{K}^{0}\pi^- \mu^+\nu_{\mu}$ can be expressed in terms of five kinematic variables:
the squared invariant mass of $\bar{K}^0\pi^-$ system ($m_{\bar{K}^0\pi^-}^2$), the squared transfer momentum of $\mu^+$ and $\nu_{\mu}$ ($q^2$),
the angle between $\pi^-$ and $D^0$ direction in the $\bar{K}^0\pi^-$ rest frame ($\theta_{\bar{K}^0}$), the angle between $\nu_{\mu}$ and $D^0$ direction in the $\mu^+\nu_{\mu}$ rest frame ($\theta_{\mu}$), and the acoplanarity angle ($\chi$) between the two decay planes of $\bar{K}^0\pi^-$ and $\mu^+\nu_{\mu}$.
The differential decay rate is expressed as~\cite{prd46_5040}
\begin{eqnarray}
d^5\Gamma&=&\frac{G^2_F|V_{cs}|^2}{(4\pi)^6m^3_{D^0}}X\beta \beta_{\mu}\mathcal{I}(m_{\bar{K}^0\pi^-}^2, q^2, \theta_{\bar{K}^0}, \theta_{\mu}, \chi) \nonumber \\
&\times& dm_{\bar{K}^0\pi^-}^2dq^2d{\rm cos}\theta_{\bar{K}^0}d{\rm cos}\theta_{\mu}d\chi,
\label{eq:differential}
\end{eqnarray}
where $X=p_{\bar{K}^{0}\pi^-}m_{D^0}$, $\beta=2p^{*}/m_{\bar{K}^{0}\pi^-}$, $\beta_{\mu}=(1-m^2_{\mu}/q^2)$. Furthermore, $p_{\bar{K}^{0}\pi^-}$ is the momentum of the $\bar{K}^{0}\pi^-$ system in the $D^0$ rest frame, $p^*$ is the momentum of $\bar{K}^{0}$ in the $\bar{K}^{0}\pi^-$ rest frame, $m_{\mu}$ ($m_{D^0}$) is the known $\mu$ ($D^0$) mass~\cite{pdg24}, and $G_F$ is the Fermi coupling constant.

To extract the FF parameters, an unbinned five-dimensional maximum likelihood fit to the distributions of $m_{\bar{K}^0\pi^-}$, $q^2$, $\cos\theta_{\bar{K}^0}$, $\cos\theta_{\mu^+}$, and $\chi$ for the signal candidates within $-0.02<U_{\rm miss}<0.02$~GeV is performed in the same manner as in Ref.~\cite{D0Kspiev}.
The parameterization of the $\mathcal{S}$-wave phase takes the form of $\delta_S(m)=\delta^{1/2}_{\rm BG}$~\cite{D0Kspiev}.
In the fit, the FF ratios $r_V$ and $r_2$, the mass $M_{K^{*}(892)^-}$ and width $\Gamma_{K^{*}(892)^-}$, the relative intensity $r_S$, scattering length $a^{1/2}_{\rm S,BG}$, and the dimensionless coefficient $r_S^{(1)}$ are set to be free variables, while the effective range $b^{1/2}_{\rm S,BG}$ is fixed to $-0.81$~(GeV/$c$)$^{-1}$, taken from Ref.~\cite{prd94_032001}. 
The projected distributions of the fit onto the fitted variables are shown in Fig.~\ref{fig:formfactor}\,(b-f). The fit results are summarized in Table~\ref{tab:FitResults}. 
The goodness of the fit is estimated as in Ref.~\cite{D0Kspiev} and $\chi^2/{ndof}$ is evaluated to be 115.9/101. 

\begin{table}
\begin{center}
\caption{The fit results, where the first uncertainty is statistical and the second is systematic. } \normalsize
\begin{tabular}
{lc} \hline\hline  Variable~~~~~~~~~~~~~~~~~~~~~~~     &   Value    \\ \hline
$M_{K^{*}(892)^-}$ (MeV/$c^2$)                                      &   $892.7\pm0.5\pm0.1$          \\
$\Gamma_{K^{*}(892)^-}$ (MeV)                                      &  ~$45.6\pm0.8\pm0.1$          \\
$r_S$ (GeV)$^{-1}$                                                          & $-12.44\pm2.40\pm3.04$       \\
$a^{1/2}_{\rm S,BG}$ (GeV/$c$)$^{-1}$                          & ~~~$3.33\pm0.53\pm1.06$     \\
$r_S^{(1)}$                                                       &  ~~~$0.06\pm0.20\pm0.26$       \\
$r_V$                                                               &  ~~~$1.46\pm0.11\pm0.04$     \\
$r_2$                                                               & ~~~$0.71\pm0.08\pm0.03$     \\
\hline\hline 
\end{tabular}
\label{tab:FitResults}
\end{center}
\end{table}

The fit fraction of each component can be determined by the ratio of the decay intensity of the specific component to that of the total intensity. The fractions of $\mathcal{S}$-wave and $\mathcal{P}$-wave ($K^{*}(892)^-$) are 
$f_{S-{\rm wave}}=(5.35\pm0.87_{\rm stat})\%$ and $f_{K^{*}(892)^-}=(94.60\pm0.87_{\rm stat})\%$, respectively, where the uncertainty propagation includes correlations among the underlying parameters.  
The systematic uncertainties of the fit parameters and the fractions of $\mathcal{S}$-wave and $K^{*}(892)^-$ components
are defined as the difference between the nominal fit and the alternative fits with varied conditions.
The systematic uncertainties due to the requirements on $E_{\gamma\,{\rm \max}}$, $M_{\bar{K}^0\pi^-\mu^+(\pi^0)}$, $\mu_{\rm PID}$, background subtraction, tracking and PID, the possible 
$\mathcal{D}$-wave component and the fixed $b^{1/2}_{\rm S,BG}$ are evaluated in the same way as in Ref.~\cite{D0Kspiev}. The total absolute systematic uncertainty for each fit parameter is also listed in Table~\ref{tab:FitResults}.

The value of $A_1(0)$ is obtained by integrating Eq.~(\ref{eq:differential}), restricted to the $K^*(892)^{-}$ contribution, over the three angles:
\begin{equation}
\small
\frac{d^2\Gamma}{dq^2dm^2_{\bar{K}^0\pi^-}}=\frac{2X_{\mu}}{9}\frac{G^2_F|V_{cs}|^2}{(4\pi)^5m^2_{D^0}}p_{\bar{K}^0\pi}\beta (|\mathcal{F}_{11}|^2+|\mathcal{F}_{21}|^2+|\mathcal{F}_{31}|^2 ),
\label{eq:inte}
\end{equation}
where $X_{\mu}=(1-m^2_{\mu}/q^2)^2[1+m^2_{\mu}/(2q^2)]$. The formulas for the FFs $\mathcal{F}_{11,21,31}$ are given in Ref.~\cite{D0Kspiev}.
The two-dimensional integration over $m^2_{\bar{K}^0\pi^-}$ and $q^2$ in Eq.~(\ref{eq:inte}) is constrained by the decay width $\Gamma(D^0\rightarrow K^{*}(892)^-\mu^+\nu_{\mu})$ which is derived by the measured $\mathcal{B}(D^0\rightarrow K^{*}(892)^-\mu^+\nu_{\mu})$ and the $D^0$ lifetime 
$\tau_{D^0}=410.3\pm1.0~{\rm fs}$~\cite{pdg24}. We then obtain $A_1(0)=0.623\pm0.008_{\rm stat}\pm0.008_{\rm syst}$, which is measured for the first time using the decay $D^0\rightarrow K^{*}(892)^-\mu^+\nu_{\mu}$. 
 
In summary, with $7.9~\mathrm{fb}^{-1}$ of $e^+e^-$ annihilation data collected at $\sqrt{s}=3.773$ GeV by the BESIII detector, the absolute BF of $D^0\rightarrow \bar{K}^0\pi^-\mu^+\nu_{\mu}$ is measured for the first time to be $\mathcal{B}(D^0\rightarrow \bar{K}^0\pi^-\mu^+\nu_{\mu})= (1.373 \pm 0.020_{\rm stat} \pm 0.023_{\rm syst})\%$.
By analyzing the dynamics of the $D^0\rightarrow \bar{K}^0\pi^-\mu^+\nu_{\mu}$ decay, the $\mathcal{S}$-wave component is measured with a fraction $f_{S-{\rm wave}} = (5.35\pm 0.87_{\rm stat} \pm 0.71_{\rm syst})\%$, leading to $\mathcal{B}[D^0\rightarrow (\bar{K}^0\pi^-)_{S-{\rm wave}}\mu^+\nu_{\mu}] = (0.073 \pm 0.012_{\rm stat} \pm 0.010_{\rm syst})\%$.
The $\mathcal{P}$-wave component is observed with a fraction of $f_{K^{*}(892)^-}=(94.60 \pm 0.87_{\rm stat} \pm 0.71_{\rm syst})\%$ and the corresponding BF is given as $\mathcal{B}(D^0\rightarrow K^*(892)^-\mu^+\nu_{\mu}) = (1.948 \pm 0.033_{\rm stat} \pm 0.036_{\rm syst})\%$.
The FF ratios in the $D^0\rightarrow K^{*}(892)^-\mu^+\nu_{\mu}$ decay are determined to be
$r_V = 1.46 \pm 0.11_{\rm stat} \pm 0.04_{\rm syst}$ and
$r_2 = 0.71 \pm 0.08_{\rm stat} \pm 0.03_{\rm syst}$. In the meanwhile, using the FF parameters measured in this work, we also determine the ratio of partial widths for longitudinal and transverse $K^*$ polarizations as defined in Refs.~\cite{PRD91_094009,EPJC77_587} to be $R_{L/T}=\Gamma_L/\Gamma_T=1.21\pm0.05_{\rm stat+syst}$, and the ratio of partial widths for positive and negative helicity defined in Ref.~\cite{EPJC77_587} to be $R_{+/-}=\Gamma_+/\Gamma_-=0.28\pm0.03_{\rm stat+syst}$ for the decay $D^0\rightarrow K^{*}(892)^-\mu^+\nu_{\mu}$ for the first time. These results provide important information in evaluating the pseudoscalar Wilson coefficient $c_P^{(\ell)}$ as in Ref.~\cite{PRD91_094009}.

\begin{figure}[tp!]
\begin{center}
   \begin{minipage}[t]{4.2cm}
   \includegraphics[width=\linewidth]{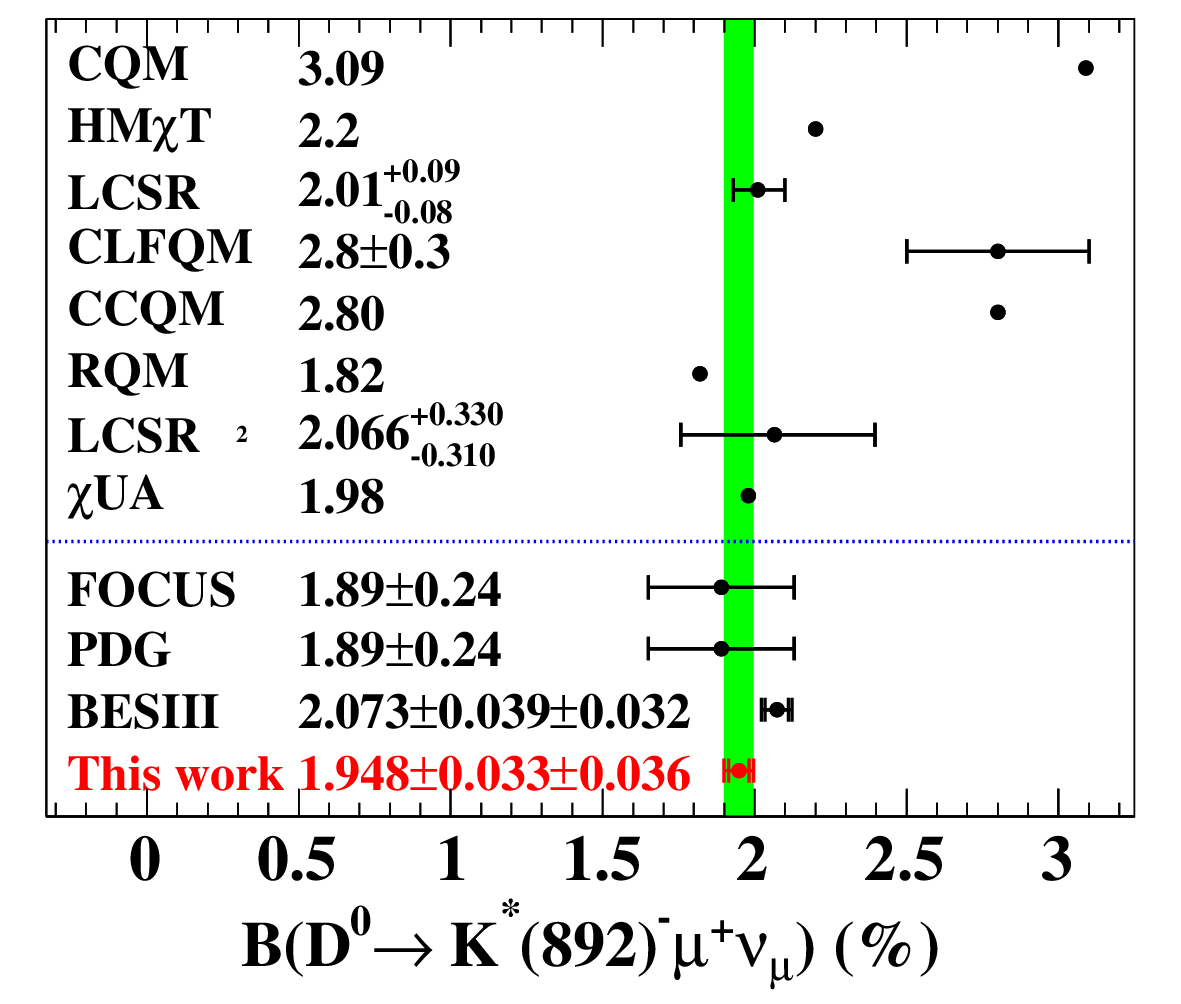}
   \end{minipage}
   \begin{minipage}[t]{4.2cm}
   \includegraphics[width=\linewidth]{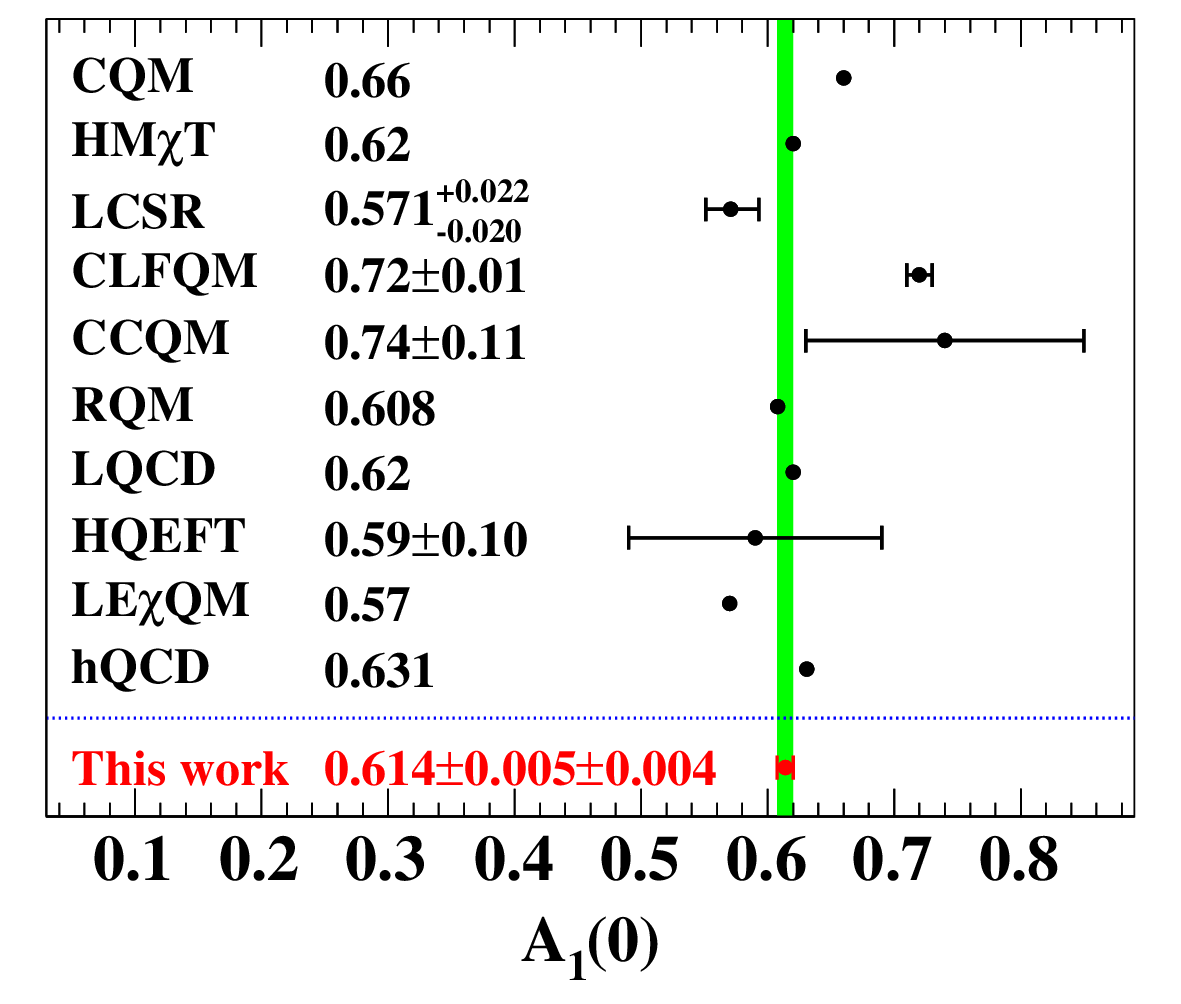}
   \end{minipage}   
   \caption{(Color online)~Comparisons of the measured $\mathcal{B}(D^0\rightarrow K^{*}(892)^-\mu^+\nu_{\mu})$ and $A_1(0)$ with various theoretical calculations from the Lattice QCD~\cite{Lattice}, LCSR~\cite{IJMPA21_6125}, LCSR~2~\cite{PRR2}, CQM~\cite{PRD62_014006,PRD96_016017}, HQEFT~\cite{PRD67_014024}, HM$\chi$T~\cite{PRD72_034029}, CLFQM~\cite{JPG39_025005,EPJC77_587}, LE$\chi$QM~\cite{PRD89_034013}, CCQM~\cite{FrontPhys14_64401}, RQM~\cite{prd101_013004}, hQCD~\cite{prd109_026008}, $\chi$UA~\cite{PRD92_054038}, and measurements from FOCUS~\cite{plb607_67}, BESIII~\cite{2403.10877}, and PDG averaged results~\cite{pdg24}. }
\label{fig:cmpbf}
\end{center}
\end{figure}

\begin{figure}[tp!]
\begin{center}
   \begin{minipage}[t]{6cm}
   \includegraphics[width=\linewidth]{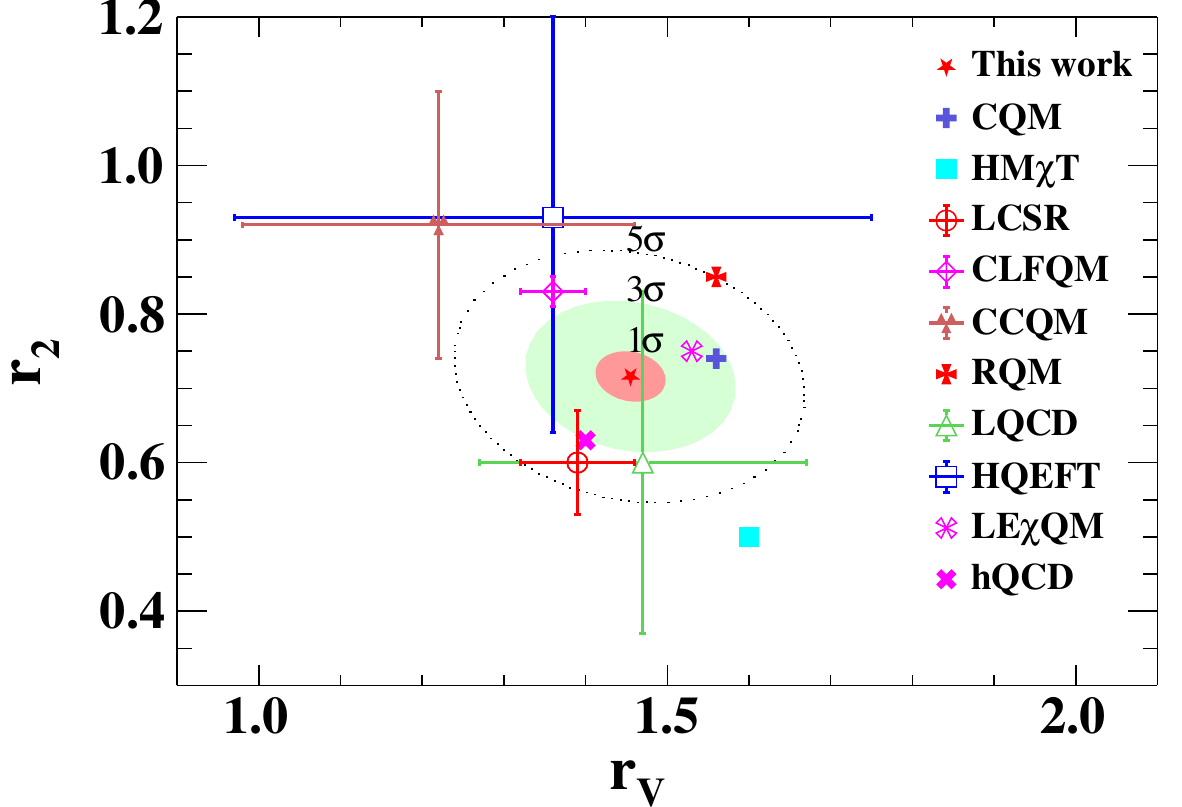}
   \end{minipage}    
   \caption{(Color online)~Comparisons of the measured $r_V$ and $r_2$ in this work with various theoretical calculations~\cite{Lattice,IJMPA21_6125,PRR2,PRD62_014006,PRD96_016017,PRD67_014024,PRD72_034029,JPG39_025005,EPJC77_587,PRD89_034013,FrontPhys14_64401,prd101_013004,prd109_026008}. The correlation coefficients between the measured $r_V$ and $r_2$ is $-0.17$. }
\label{fig:cmpformfactor}
\end{center}
\end{figure}

Averaging the results of $r_{V}$, $r_{2}$ and $A_1(0)$ reported previously in Ref.~\cite{D0Kspiev,2403.10877},
we obtain the most precise measurements of FF parameters $r_{V}=1.456\pm0.040_{\rm stat}\pm0.016_{\rm syst}$, $r_{2}=0.715\pm0.031_{\rm stat}\pm0.014_{\rm stat}$, and $A_1(0)=0.614\pm0.005_{\rm stat}\pm0.004_{\rm syst}$ in the $D^0\rightarrow K^*(892)^-$ transition. The comparisons of the $\mathcal{B}(D^0\rightarrow K^*(892)^-\mu^+\nu_{\mu})$, $A_1(0)$, $r_{V}$, and $r_{2}$ between this measurement and theoretical calculations~\cite{Lattice,IJMPA21_6125,PRR2,PRD62_014006,PRD96_016017,PRD67_014024,PRD72_034029,JPG39_025005,EPJC77_587,PRD89_034013,FrontPhys14_64401,prd101_013004,prd109_026008} are shown in Figs.~\ref{fig:cmpbf} and~\ref{fig:cmpformfactor}.
At a confidence level (C.L.) of 95\%, our measured BF disfavors the central values calculated from the CQM~\cite{PRD62_014006,PRD96_016017}, HM$\chi$T~\cite{PRD72_034029}, CLFQM~\cite{JPG39_025005,EPJC77_587}, and CCQM~\cite{FrontPhys14_64401}. While the measured $r_V$ and $r_2$ in this work disfavor the central values calculated from the HQEFT~\cite{PRD67_014024}, HM$\chi$T~\cite{PRD72_034029}, CLFQM~\cite{JPG39_025005,EPJC77_587}, CCQM~\cite{FrontPhys14_64401}, and RQM~\cite{prd101_013004} by more than 3 standard deviations. Furthermore, using the $\mathcal{B}(D^0\rightarrow K^{*}(892)^-e^+\nu_{e})$ and $\mathcal{B}(D^0\rightarrow \bar{K}^0\pi^-e^+\nu_{e})$ measured in Ref.~\cite{D0Kspiev}, we obtain the relative ratios between $\mu$ and $e$ channels to be:
\begin{eqnarray}
\mathcal{R}^{\mu/e}_{K^*(892)}&=&0.955\pm0.022_{\rm stat}\pm0.017_{\rm syst}, \nonumber \\
\mathcal{R}^{\mu/e}_{\bar{K}^0\pi}&=&\frac{\mathcal{B}(D^0\rightarrow \bar{K}^0\pi^-\mu^+\nu_{\mu})}{\mathcal{B}(D^0\rightarrow \bar{K}^0\pi^-e^+\nu_{e})} \nonumber \\ 
&=&0.951\pm0.020_{\rm stat}\pm0.016_{\rm syst}.~~~~~~~~~~ \nonumber 
\end{eqnarray}
These ratios are in good agreement with the calculations in Refs.~\cite{IJMPA21_6125,PRD96_016017,EPJC77_587,FrontPhys14_64401,PRD89_034013,prd101_013004,PRD92_054038}, but disfavour the calculation in Ref.~\cite{PRR2} at 68\% C.L.  The results presented in this Letter provide the most powerful tests and constraints on various theoretical calculations especially for QCD theory, and play an important role in understanding
the dynamics of SL decays of the charmed hadrons in the non-perturbative region.

\acknowledgments
The BESIII Collaboration thanks the staff of BEPCII (https://cstr.cn/31109.02.BEPC) and the IHEP computing center for their strong support. This work is supported in part by National Key R\&D Program of China under Contracts Nos. 2020YFA0406400, 2020YFA0406300, 2023YFA1606000, 2023YFA1606704; National Natural Science Foundation of China (NSFC) under Contracts Nos. 11635010, 11935015, 11935016, 11935018, 12022510, 12025502, 12035009, 12035013, 12061131003, 12192260, 12192261, 12192262, 12192263, 12192264, 12192265, 12221005, 12225509, 12235017, 12375090, 12475092, 12361141819; the Chinese Academy of Sciences (CAS) Large-Scale Scientific Facility Program; CAS under Contract No. YSBR-101; 100 Talents Program of CAS; The Institute of Nuclear and Particle Physics (INPAC) and Shanghai Key Laboratory for Particle Physics and Cosmology; Agencia Nacional de Investigación y Desarrollo de Chile (ANID), Chile under Contract No. ANID PIA/APOYO AFB230003; German Research Foundation DFG under Contract No. FOR5327; Istituto Nazionale di Fisica Nucleare, Italy; Knut and Alice Wallenberg Foundation under Contracts Nos. 2021.0174, 2021.0299; Ministry of Development of Turkey under Contract No. DPT2006K-120470; National Research Foundation of Korea under Contract No. NRF-2022R1A2C1092335; National Science and Technology fund of Mongolia; National Science Research and Innovation Fund (NSRF) via the Program Management Unit for Human Resources \& Institutional Development, Research and Innovation of Thailand under Contract No. B50G670107; Polish National Science Centre under Contract No. 2024/53/B/ST2/00975; Swedish Research Council under Contract No. 2019.04595; U. S. Department of Energy under Contract No. DE-FG02-05ER41374. This paper is also supported by the Fundamental Research Funds for the Central Universities, and the Research Funds of Renmin University of China under Contract No. 24XNKJ05.



\begin{thebibliography}{99}
\bibitem{BFK0} The BFs of $K^0\rightarrow K^0_S$ and $K^0_S\rightarrow \pi^+\pi^-$ are not included in the detection efficiency. 
The $K^0_S$ fraction of a $K^0$ is taken as 50\% and the BF of $K^0_S\rightarrow \pi^+\pi^-$ is taken from PDG~\cite{pdg24}.
\end{thebibliography}
\end{document}